\documentclass{desyproc}

\begin{document}
\title{SM*A*S*H 
}

\author{{\slshape  Andreas Ringwald,}\\[1ex]
Deutsches Elektronen-Synchrotron (DESY), Hamburg, Germany}

\contribID{familyname\_firstname}

\confID{13889}  
\desyproc{DESY-PROC-2016-XX}
\acronym{Patras 2016} 
\doi  

\maketitle

\begin{abstract}
We present a minimal model for particle physics and cosmology. The Standard Model (SM) 
particle content is extended by three right-handed SM-singlet neutrinos $N_i$ and a vector-like quark $Q$, all of them being charged under a global lepton number and Peccei-Quinn (PQ) $U(1)$ symmetry  which is  spontaneously broken by the vacuum expectation value $v_\sigma \sim 10^{11}$ GeV of a SM-singlet complex scalar field $\sigma$. Five fundamental problems 
-- neutrino oscillations, baryogenesis, dark matter, inflation, strong CP problem -- are solved at one stroke in this 
model, dubbed ``SM*A*S*H" (Standard Model*Axion*Seesaw*Higgs portal inflation). 
 It can be probed decisively by upcoming cosmic microwave
background and axion dark matter experiments.
\end{abstract}

\section{The quest for a minimal model of particle cosmology}

The discovery of the Higgs boson has marked the completion of the SM particle content. 
However, observations in particle physics, astrophysics, and cosmology point to the existence of 
particles and interactions beyond the SM. In fact, the SM lacks an explanation of 
{\em i)} neutrino oscillations,
{\em ii)} the baryon asymmetry of the Universe,
{\em iii)} dark matter,
{\em iv)} inflation, and 
{\em v)} the non-observation of strong CP violation.

Remarkably, problems {\em 1)-3)} are solved in 
the Neutrino Minimal SM ($\nu$MSM)~\cite{Asaka:2005an,Asaka:2005pn}: a minimal extension of the SM by three right-handed singlet neutrinos
$N_i$,
having Dirac masses $m_D=F v/\sqrt{2}$ arising from Yukawa couplings $F$ with the Higgs ($H$) and lepton ($L_i$)
doublets, as well as explicit Majorana masses $M$, 
\begin{equation}
  {\cal L}\supset -\big[ F_{ij}L_i\epsilon H N_j+\frac{1}{2}M_{ij} N_i  N_j \big]\,,
\label{yuk_nuMSM}
\end{equation}
(in Weyl spinor notation).  
In the seesaw limit, $M\gg m_D$, the neutrino mass spectrum splits into a light set given by the eigenvalues 
$m_1<m_2<m_3$ of 
the matrix $m_\nu = - m_D M^{-1} m_D^T$, with the eigenstates corresponding mainly to mixings of the active 
left-handed neutrinos $\nu_\alpha$, and a heavy set given by the eigenvalues $M_1<M_2<M_3$ of the matrix $M$, with the eigenstates corresponding to mixings of the sterile right-handed neutrinos $N_i$. Problem {\em 1)} is thus solved by the usual seesaw type-I mechanism. Intriguingly, problems {\em 2)} and {\em 3)} can be solved simultaneously if 
 $M_1\sim $\,keV and  $M_2\sim M_3\sim $\,GeV. 
In fact, in this case $N_{2,3}$ create flavored lepton asymmetries from CP-violating oscillations in the early Universe
which are crucial for the generation of the baryon asymmetry of the Universe via flavored leptogenesis and of the lightest sterile neutrino $N_1$ -- the dark matter candidate of the
$\nu$MSM -- by the MSW effect. Moreover, it was argued in Ref.  \cite{Bezrukov:2007ep} that also problem {\em 4)} can be solved in the $\nu$MSM by allowing a non-minimal coupling of the Higgs field to the Ricci scalar,  
$S\supset - \int d^4x\sqrt{- g}\,\xi_H\, H^\dagger H  R$, which promotes the Higgs field to an inflaton candidate. 

However, the success of the $\nu$MSM as a minimal model of particle cosmology is threatened by several facts. 
First of all, recent findings in astrophysics have seriously constrained the parameter space for $N_1$ as a dark matter candidate \cite{Schneider:2016uqi,Perez:2016tcq}. Secondly, the large value of the non-minimal coupling 
$\xi_H \sim 10^5 \sqrt{\lambda_H}$, where $\lambda_H$ is the Higgs self-coupling, required to fit the amplitude of the scalar perturbations inferred from the 
cosmic microwave background (CMB) temperature fluctuations, imply that perturbative unitarity breaks down at the scale
$M_P/\xi_H\sim 10^{14}$\,GeV, well below the scale of inflation, $M_P/\sqrt{\xi_H}\sim 10^{16}$\,GeV,  
 making the inflationary predictions unreliable \cite{Barbon:2009ya,Burgess:2009ea}.  
Thirdly, Higgs inflation cannot be realised at all if the Higgs quartic coupling $\lambda_H$ runs negative at large (Planckian) field values due to the corrections from top quark loops. 
Although, given the current experimental uncertainties, a definite conclusion cannot yet be drawn, see e.g.~\cite{Buttazzo:2013uya,Bednyakov:2015sca}, the presently favoured central values of the strong gauge coupling and the Higgs and top quark masses imply that $\lambda_H$ becomes negative at a field value corresponding to an energy scale $\Lambda_I \sim 10^{11}$ GeV, much lower than what is required for Higgs inflation, and is thus inconsistent with it.

These three obstacles of the $\nu$MSM are circumvented in SMASH - an extension of the SM which features the Axion, 
the type-I Seesaw and Higgs portal inflation \cite{Ballesteros:2016euj,Ballesteros:2016xej} - as we will review in these proceedings.

\section{\label{model}The SMASH model}

The SM particle content is extended not only by three right-handed singlet neutrinos $N_i$, but also by a vector-like color-triplet quark $Q$, as in the KSVZ \cite{Kim:1979if,Shifman:1979if} model. The SM quarks and leptons as well as  the $N_i$ and $Q$  are  assumed to be charged under a global lepton number and PQ $U(1)$ symmetry \cite{Peccei:1977hh} which is  spontaneously broken by the vacuum expectation value $v_\sigma \sim 10^{11}\,{\rm GeV}$ of a 
SM-singlet complex scalar field $\sigma$. The  most general scalar potential  reads
\begin{equation}
\label{scalar_potential} 
\nonumber
V(H,\sigma )  = \lambda_H \left( H^\dagger H - \frac{v^2}{2}\right)^2
+\lambda_\sigma \left( |\sigma |^2 - \frac{v_{\sigma}^2}{2}\right)^2 +
2\lambda_{H\sigma} \left( H^\dagger H - \frac{v^2}{2}\right) \left( |\sigma |^2 - \frac{v_{\sigma}^2}{2}\right) ,
\end{equation}
while the most general Yukawa couplings of the new fields are given by
\begin{equation}
\nonumber
  {\cal L}\supset -\Bigg[F_{ij}L_i\epsilon H N_j+\frac{1}{2}Y_{ij}\sigma N_i  N_j 
+y\, \tilde Q \sigma Q+\,{y_{Q_d}}_{i}\sigma Q d_i+h.c.\Bigg]\,.
\label{lyukseesaw}
\end{equation}

After $U(1)$ symmetry breaking the sterile neutrinos $N_i$, the particle excitation $\rho$ of the modulus of the $\sigma$ field, and the exotic quark $Q$  get large masses $\propto v_\sigma\gg v =246$\,GeV: 
$M_{ij} = \frac{Y_{ij}}{\sqrt{2}} v_\sigma + \mathcal{O}\left(  \frac{v}{v_{\sigma}}\right)$, 
$m_\rho =   \sqrt{2\,\lambda_\sigma}\, v_{\sigma} + \mathcal{O}\left(  \frac{v}{v_{\sigma}}\right)$, and
$m_Q = \frac{y}{\sqrt{2}}\, v_{\sigma} + \mathcal{O}\left(  \frac{v}{v_{\sigma}}\right)$. 
Therefore, as far as physics around the electroweak scale or below is concerned, these heavy particles can be integrated out (unless one considers tiny Yukawa and self couplings). The corresponding low-energy Lagrangian of SMASH is identical to that of the SM, augmented by seesaw-generated neutrino masses, $m_{\nu} =  0.04\,{\rm eV}  \left( \frac{10^{11}\,{\rm GeV}}{v_{\sigma}} \right)
\left( \frac{-  F\,Y^{-1}\,F^T}{10^{-4}}\right)$, and mixing (thus solving problem {\em 1)}),  
plus one new particle: the particle excitation $A$ of the angular degree of freedom of the complex $\sigma$ field -- the Nambu-Goldstone boson of the spontaneous symmetry breaking of the $U(1)$, which is dubbed ``axion" in the literature dealing with the PQ solution of the strong CP problem \cite{Weinberg:1977ma,Wilczek:1977pj} and ``majoron" in the literature dealing with the spontaneous breaking of a global lepton symmetry. Integrating out the 
exotic quark induces an anomalous coupling of the axion field to the topological charge density in QCD, 
${\mathcal L}\supset 
- \frac{\alpha_s}{8\pi}\,\frac{A}{f_A}\,G_{\mu\nu}^c {\tilde G}^{c,\mu\nu}$, promoting the axion field to 
a dynamical theta parameter, $\theta (x)=A(x)/f_A$, which relaxes to zero in the vacuum, 
$\langle \theta\rangle =0$, thereby solving problem {\em 5)}. 
While the strong CP problem is solved for any
value of the axion decay constant $f_A=v_\sigma$, the dark matter will be comprised by axions only if $f_A$ is around $10^{11}$\,GeV, as we will see later. In this case, the axion mass is predicted to be around $m_A= 
57.0(7)\,   \left(\frac{10^{11}\rm GeV}{f_A}\right)\mu \text{eV}$ \cite{Weinberg:1977ma,Borsanyi:2016ksw}.

\begin{figure}[t]
\begin{center}
\includegraphics[width=0.32\textwidth]{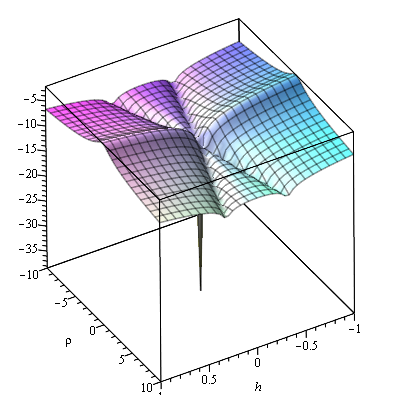}
\hfill
\includegraphics[width=0.32\textwidth]{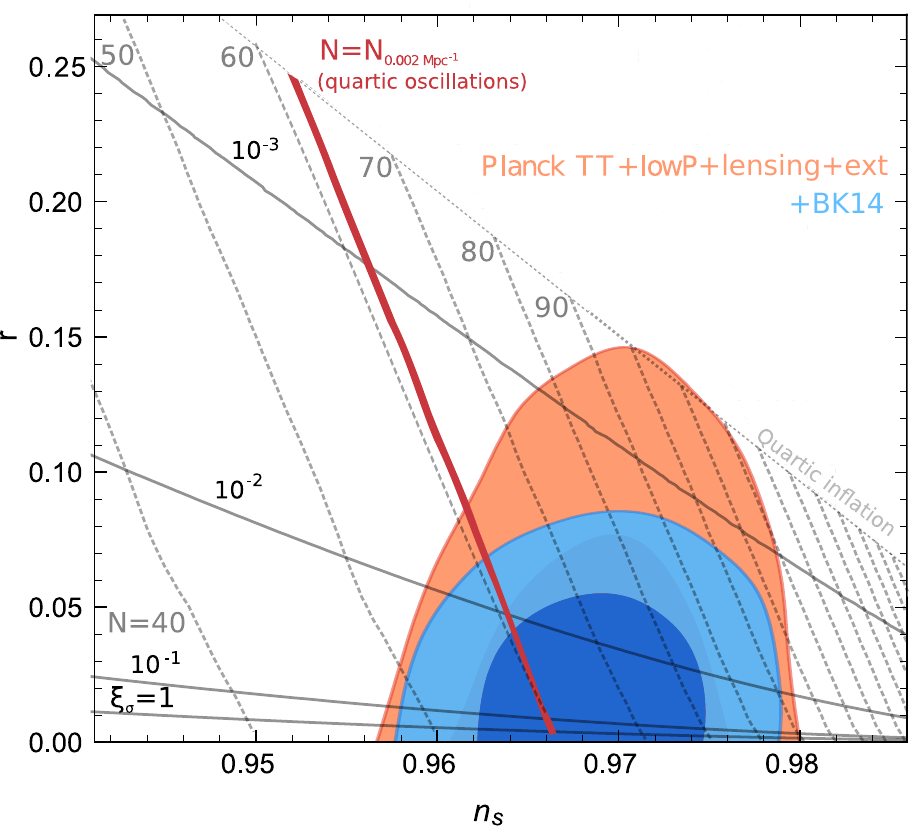}
\hfill
\includegraphics[width=0.32\textwidth]{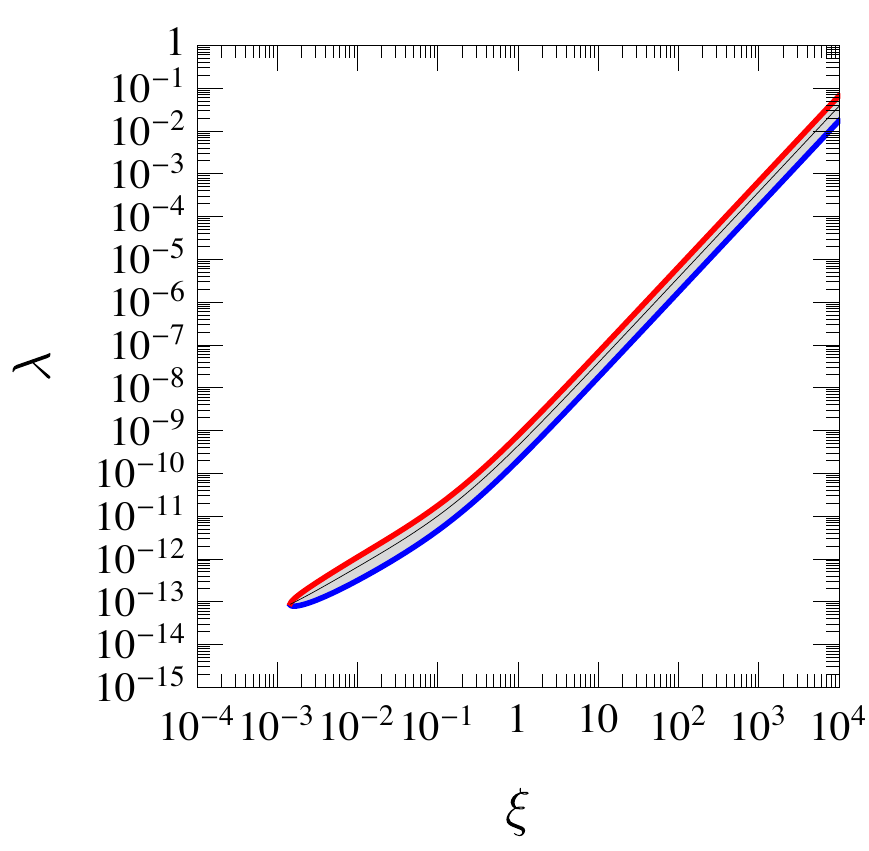}
\caption{Left: Decadic log of the SMASH scalar potential in the Einstein frame, as a function of Higgs in the 
unitary gauge, $h$, and the modulus  of $\sigma$, $\rho$, all in units of $M_P$,
for $\kappa_H<0$, $\kappa_\sigma <0$, supporting mixed Higgs-Hidden-Scalar Inflation  along one of the valleys. 
Middle: Bounds on $r$ vs.  $n_s$ \cite{Ade:2015xua}, compared to the predictions from (H)HSI in SMASH for fixed values of the non-minimal coupling $\xi_\sigma$ and the number of e-folds $N$, respectively.
Right: Required self-coupling versus non-minimal coupling to reproduce CMB results on inflation. All figures from \cite{Ballesteros:2016xej}.
}
\label{fig:inflation}       
\end{center}
\end{figure}

\section{Inflation}

The non-minimal couplings in SMASH, $  S\supset - \int d^4x\sqrt{- g}\,\left[
    \xi_H\, H^\dagger H+\xi_\sigma\, \sigma^* \sigma  
  \right] R$,  stretch the scalar potential in the 
Einstein frame, which makes it convex and asymptotically flat at large field values. 
Depending on the signs of the parameters $\kappa_H  \equiv \lambda_{H\sigma} \xi_H - \lambda_H \xi_\sigma$ and $\kappa_\sigma  \equiv \lambda_{H\sigma} \xi_\sigma - \lambda_\sigma \xi_H$,  it can support Higgs Inflation (HI),  Hidden Scalar Inflation  
(HSI), or even mixed Higgs-Hidden Scalar Inflation  (HHSI)   (cf. Fig.\  \ref{fig:inflation} (left)).  
For $\xi\simeq 10^5\sqrt{\lambda}\gtrsim 10^{-3}$,   where
\begin{equation}
{\xi} \equiv 
\left\{
\begin{array}{ll} 
\xi_H
,  & \mathrm{for\ HI},  \\
\xi_\sigma
,  & \mathrm{for\ HSI},  \\ 
\xi_\sigma 
,  & \mathrm{for\ HHSI},
\end{array}
\right.
\hspace{6ex}
{\lambda} \equiv 
\left\{
\begin{array}{ll} 
\lambda_H
,  & \mathrm{for\ HI},  \\
\lambda_\sigma
,  & \mathrm{for\ HSI},  \\ 
\lambda_\sigma \left( 1-\frac{\lambda_{H\sigma}^2}{\lambda_\sigma\lambda_H} \right)
,  & \mathrm{for\ HHSI},
\end{array}
\right.
\end{equation}
the predicted values of the CMB observables such as  the amplitude of scalar perturbations $A_s$,  the spectral 
index $n_s$, and the 
tensor-to-scalar ratio $r$ are in perfect consistency with the current observations, 
see e.g. {Fig.\ \ref{fig:inflation} (middle).} Importantly, for (H)HSI, the effective self-coupling $\lambda$
is a free parameter and therefore can be chosen small, 
$\lambda \sim 10^{-10}$, 
such that the required non-minimal coupling 
to fit the amplitude of primordial scalar perturbations is of order unity, $\xi_\sigma \sim 1$, cf. Fig.\ \ref{fig:inflation} (right). 
In this region of parameter space, the perturbative predictivity of SMASH is guaranteed and superior to HI, which necessarily {operates} at large $\xi_H$, since $\lambda_H$ is not small. 
Remarkably, the requirement of predictive inflation, free of unitarity problems, demands $r\gtrsim 0.01$, which will be probed by CMB experiments such as LiteBIRD and PRISM.

\section{Stability}

Self-consistency of inflation in SMASH requires a positive scalar potential all the way up to the Planck scale.  
Importantly, the Higgs portal term $\propto \lambda_{H\sigma}$ in the scalar potential helps to ensure absolute stability in the Higgs direction via the threshold stabilisation mechanism pointed out 
in~\cite{Lebedev:2012zw,EliasMiro:2012ay}.
We have found that stability can be achieved if the threshold parameter $\delta = \lambda_{H\sigma}^2/\lambda_\sigma$ is between $10^{-3}$ and $10^{-1}$. Instabilities could also originate  in the $\sigma$ direction, due to quantum corrections from the right-handed neutrinos $N_i$ and the exotic quark $Q$. Stability in the 
$\sigma$ direction then restricts their Yukawas to   $\sum Y^4_{ii}+6 y^4\lesssim 16\pi^2\lambda_\sigma/\log\left(30 M_P/\sqrt{2\lambda_\sigma}v_\sigma \right)$.

\begin{figure}[t]
\begin{center}
\includegraphics[width=0.7\textwidth]{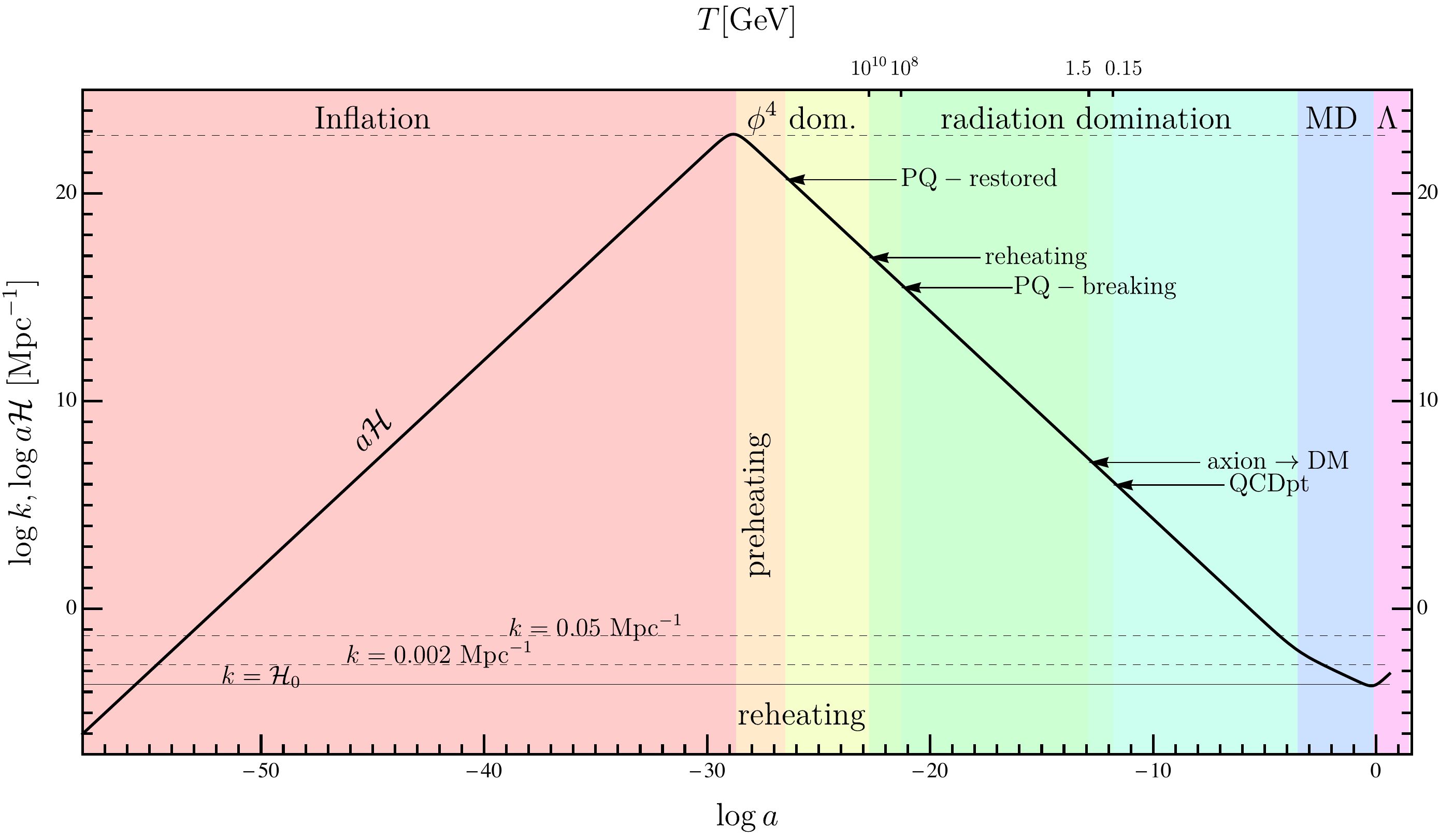}
\caption{\small The history of the Universe in SMASH HHSI, emphasising the transition from inflation to radiation-domination-like Universe expansion $a{\cal H}\propto 1/a$ before standard matter and cosmological constant domination epochs \cite{Ballesteros:2016xej}.}  
\label{fig:history}       
\end{center}
\end{figure}

\section{Reheating}

Both in (H)HSI, slow-roll inflation ends at a value of $\rho\sim \mathcal{O}(M_P)$, where the effect of
$\xi_\sigma\sim 1$ is negligible and the inflaton starts to undergo Hubble-damped oscillations in a quartic potential, with the Universe expanding as in a radiation-dominated era, which lasts until reheating, cf. Fig. \ref{fig:history}. After the latter, radiation domination continues, though driven by a bath of relativistic particles. This fixes the thick red line in Fig.\ \ref{fig:inflation} (middle) as the prediction for $r$, $n_s$ and $N$ in SMASH.
The fluctuations of $\sigma$ grow fast due to parametric resonance while the inflaton background oscillates in its
quartic potential, leading to a rapid restoration of the PQ symmetry after about 14 oscillations. 
The following reheating stage differs considerably for HSI and HHSI. In the former, the large induced particle masses
quench inflaton decays or annihilations into SM particles, resulting in a low reheating temperature, 
$T_R\sim 10^7$\,GeV, 
such that the produced relativistic axions are never thermalized. 
Correspondingly, HSI predicts a significant amount of cosmic axion background radiation  (CAB): an increase 
$\Delta N_\nu^{\rm eff}={\mathcal O}(1)$
of the effective number of relativistic neutrino species beyond the SM value
$N_\nu^{\rm eff}({\rm SM})=3.046$ \cite{Mangano:2001iu}. This disvafors HSI, since the current results from CMB and
baryon acoustic oscillations yield $N_\nu^{\rm eff} = 3.04 \pm 0.18$ at 68\% CL and thus do not allow an additional contribution of order one 
 \cite{Ade:2015xua}. 
For this reason, inflation in SMASH must be of HHSI type, and therefore the inflaton contains a (small) Higgs component.
The latter allows for efficient reheating of the Universe by the production of SM gauge bosons.
The reheating temperature in this case is predicted to be around {$T_R\sim  10^{10}$\,GeV.
Such temperature ensures a thermal restoration of the PQ symmetry for the relevant region of parameter space,  since the critical temperature $T_c$ of the PQ phase transition goes as  ${T_c}/{v_\sigma}\simeq{2\sqrt{6\lambda_\sigma}}/{\sqrt{8(\lambda_\sigma+\lambda_{H\sigma})+\sum_i Y^2_{ii}+6 y^2}}$. A thermal background of axions is produced at this stage which later decouples and results in a moderate CAB corresponding to  $\triangle N_\nu^{\rm eff}\simeq 0.03$, a prediction which may be 
checked in a future CMB polarisation experiment.}

\section{Dark Matter}

Dark matter is produced in SMASH by the re-alignment mechanism~\cite{Preskill:1982cy,Abbott:1982af,Dine:1982ah} and the decay of topological defects (axion strings and domain walls) \cite{Kawasaki:2014sqa}.
In order to account for all of the cold dark matter in the Universe, the PQ symmetry breaking scale is predicted to be in the range $3\times 10^{10}\,\mathrm{GeV}\lesssim f_A \lesssim   1.2\times 10^{11}\,\mathrm{GeV}$, corresponding to an axion mass in the range $50\,\mu\mathrm{eV}\lesssim m_A \lesssim 200 \,\mu\mathrm{eV}$ \cite{Ballesteros:2016xej,Borsanyi:2016ksw}. Here, the 
 uncertainty  originates mainly from the difficulty in predicting the relative importance of the two main production  mechanisms of axionic dark matter, i.e. re-alignment and topological defect decay.
Importantly, the axion dark matter mass window will be probed  in the upcoming decade by axion dark matter direct detection experiments such as CULTASK, MADMAX, and ORPHEUS.

\section{Baryogenesis}

The origin of the baryon asymmetry of the Universe  is explained in SMASH by thermal leptogenesis \cite{Fukugita:1986hr}. In HHSI, after reheating and thermal PQ restoration, the RH neutrinos become massive and 
at least the lightest RH neutrino $N_1$ will retain an equilibrium abundance.  However the stability bound on 
$M_1\lesssim 10^8\, (\lambda/10^{-10})^{1/4} (v_\sigma/10^{11} {\rm GeV})$\,GeV, for a hierarchical $N_i$ spectrum ($M_3=M_2=3M_1$), is just borderline 
compatible with vanilla leptogenesis from the decays of $N_1$, 
which demands $M_1\gtrsim 5\times10^8$ GeV \cite{Giudice:2003jh,Buchmuller:2004nz}. 
Nevertheless, leptogenesis can occur with a mild resonant enhancement \cite{Pilaftsis:2003gt} for a less hierarchical RH neutrino spectrum, which relaxes the stability bound and ensures that all the RH neutrinos remain in equilibrium after the phase transition.


\section{Acknowledgments}

Many thanks to Guillermo Ballesteros, Javier Redondo and Carlos Tamarit for the great collaboration.


\begin{footnotesize}

\end{footnotesize}


\end{document}